\documentclass[a4paper, USenglish]{lipicsoid-2019}

\hideLIPIcs
\usepackage{etex}
\usepackage{amsmath}
\usepackage{amssymb,amsthm,url}
\usepackage{stmaryrd}
\usepackage{todonotes}
\usepackage{comment}
\usepackage[numbers, sort&compress]{natbib} 
\usepackage{doi} 
\usepackage{bibentry}
\nobibliography*

\hypersetup{
  colorlinks=true,
  linkcolor=black,
  citecolor=black,
  filecolor=black,
  urlcolor=[rgb]{0,0.1,0.5},
  pdftitle={Bibliography of distributed approximation beyond bounded degree},
  pdfauthor={}
}

\title{Bibliography of distributed approximation beyond bounded degree}

\author{Laurent Feuilloley}%
{CNRS and University of Lyon}%
{laurent.feuilloley@univ-lyon1.fr}%
{http://orcid.org/0000-0002-3994-0898}%
{}

\authorrunning{L. Feuilloley}

\newcommand{\problem}[1]{\emph{Problem} : #1.\\}
\newcommand{\class}[1]{\emph{Graph class} : #1.\\}
\newcommand{\results}[1]{\emph{Results} : #1.\\}
\newcommand{\resultsw}[1]{\emph{Results} : #1}
\newcommand{\approxim}[1]{\emph{Approximation ratio} : #1.\\}
\newcommand{\complexity}[1]{\emph{Complexity} : #1.\\}
\newcommand{\notes}[1]{\emph{Notes} : #1\\}

\newcommand{\polylog}{\text{polylog}\,}

\begin{document}
\maketitle

\begin{abstract}
This document is an informal bibliography of the papers dealing with distributed approximation algorithms. A classic setting for such algorithms is bounded degree graphs, but there is a whole set of techniques that have been developed for other classes. These later classes are the focus of the current note. These classes have a geometric nature (planar, bounded genus and unit-disk graphs) and/or have bounded parameters (arboricity, expansion, growth, independence) or forbidden structures (forbidden minors). 
\end{abstract}

\paragraph*{Preliminary notes.}
\begin{itemize}
\item Some papers cited here are not about approximation, but are nonetheless included because of the strong connections in terms of techniques.
\item In the case of graphs that can be embedded on a surface (\emph{e.g.} planar graphs), we assume that the nodes \emph{do not have the knowledge of their locations}. Some papers, especially in the literature about robots, assume such knowledge, and are not cited here.
\item There has been series of papers improving one on the other, either generalizing to larger classes, or improving the approximation ratio. We choose to list the papers from the most recent to the oldest, to have the up-to-date results first. The bibliography generated at the end of this document is in alphabetical order to allow quick access to a specific reference. When citing a paper in the section of another paper, we first give the reference in term of section number and then as a pointer to the bibliography. Also, since conference and journal versions sometimes differ, we cite all versions.
\item This is not a very polished document, and we may have missed references, or not cited properly every paper. Please let us know if you find any mistake or omission.
\end{itemize}

\paragraph*{Graph classes considered.}

The main graphs classes considered in this document are : bounded-expansion, planar, unit-disk, bounded-genus, bounded-arboricity, bounded-independence (a.k.a. bounded-growth) and minor-free (= minor-closed) graphs. The relations between these classes are the following: 
\begin{itemize}
\item Outerplanar $\subseteq$ planar $\subseteq$ bounded-genus $\subseteq$ minor-free $\subseteq$ bounded expansion.
\item Bounded expansion $\subseteq$ bounded arboricity, and also bounded-genus $\subseteq$ bounded arboricity.
\item Unit-disk graphs $\subseteq$ bounded-independence.
\end{itemize}

\paragraph*{Versions.} The first verison of this document appeared on arxiv in January 2020 and was revised in April 2020. The current version is from October 2023.

\paragraph*{Acknowledgments.} I thank Sebastian Siebertz and Yi-Jun Chang for feedback and pointers to papers I had missed.

\newpage{}

\section{Distributed domination on sparse graph classes -
\cite{HeydtKOSV23}}
\label{sec:HeydtKOSV23}

\problem{Minimum dominating set}
\class{Bounded-expansion  and planar}
\approxim{Constant on bounded-expansion, and $11+\epsilon$ on planar}
\complexity{Constant}
\notes{This journal version subsumes Paper~\ref{sec:KublenzSV21}~\cite{KublenzSV21} and Paper~\ref{sec:HeydtSV22}~\cite{HeydtSV22}, by the same authors. 
It simplifies the arguments of Paper~\ref{sec:CzygrinowHWW18}~\cite{CzygrinowHWW18}, identifies bounded-$\nabla_1$ as the key property for the technique to work (this is actually more general than bounded expansion, but less than bounded degeneracy). The direct application of the result to planar leads to a 21-approx, but a more careful analysis leads to $11+\epsilon$. }

\section{Efficient Distributed Decomposition and Routing Algorithms in Minor-Free Networks and Their Applications - \cite{Chang23}}

\problem{Maximum independent set, in \textsc{Congest}}
\class{Minor-free}
\approxim{$1-\epsilon$}
\complexity{$\epsilon^{-1}\log^*n+poly(\epsilon^{-1})$}
\notes{The results improves on the one of~Paper~\ref{sec:ChangS22}~\cite{ChangS22} for this problem.
It gets a similar bound as Paper~\ref{sec:CHW08-disc}~\cite{CHW08-disc}, but in the \textsc{Congest} model. Previous algorithms with that complexity were in the \textsc{Local} model with very large messages. This is done via a new low-diameter decomposition.}

\section{Recurrent Problems in the LOCAL model - \cite{AgrawalAPR22}}
\label{sec:AgrawalAPR22}

\problem{Minimum Client Dominating Set (in the supported model)}
\class{Trees and planar graphs}
\approxim{1+$\epsilon$}
\complexity{Constant}
\notes{This is in a different model, the SUPPORTED model, where some preprocessing is allowed. The preprocessing allows to bypass some locality lower bounds. The paper appeared as a brief announcement in 2023~\cite{AgrawalAPR23}.}

\section{Near-Optimal Distributed Dominating Set in Bounded Arboricity Graphs - \cite{Dory0I22}}
\label{sec:Dory0I22}

\problem{Minimum weighted dominating set}
\class{Bounded arboricity $\alpha$, bounded degree $\Delta$}
\approxim{$(2\alpha+1)(1+\epsilon)$}
\complexity{$O(\epsilon^{-1}\log \Delta)$}
\notes{The result is nearly optimal (proved in the paper). Improves on Paper~\ref{sec:LenzenW10}~\cite{LenzenW10}, on Paper~\ref{sec:BansalU17}~\cite{BansalU17}, and on Paper~\ref{sec:MorganSW21}~\cite{MorganSW21}.}

\section{Narrowing the LOCAL-CONGEST Gaps in Sparse Networks via Expander Decompositions - \cite{ChangS22}}
\label{sec:ChangS22}

\problem{Maximum weighted matching and maximum independent set, in \textsc{Congest}}
\class{Minor-free}
\approxim{$1- \epsilon$}
\complexity{$\epsilon^{-O(1)} \log^{O(1)}n$ randomized for MWM, and $\epsilon^{-O(1)} \log^{O(1)}n$ randomized and $\epsilon^{-O(1)} 2^{O(\sqrt{\log n \log \log n})}$ deterministic for MIS}
\notes{Since then, the same approx/complexity for MWM has been obtained in general graphs~\cite{HuangS23}.
Other problems are also considered in the paper (correlated clustering and testing variants). The authors build a framework based on expander decomposition. }

\section{Local planar domination revisited - \cite{HeydtSV22}}
\label{sec:HeydtSV22}

\problem{Minimum dominating set}
\class{Planar graphs}
\approxim{20}
\complexity{Constant}
\notes{Improves on the constant from Paper~\ref{sec:Wawrzyniak14}~\cite{Wawrzyniak14} (and on the one of Paper~\ref{sec:AlipourFK20}~\cite{AlipourFK20} for triangle-free planar graphs). 
The algorithm is a mix of ideas from Paper~\ref{sec:LenzenPW13}~\cite{LenzenPW13}, Paper~\ref{sec:CzygrinowHWW18}\cite{CzygrinowHWW18} and Paper~\ref{sec:KublenzSV21} \cite{KublenzSV21}.}

\section{Algorithms for the Minimum Dominating Set Problem in Bounded Arboricity Graphs: Simpler, Faster, and Combinatorial - \cite{MorganSW21}}
\label{sec:MorganSW21}

\problem{Minimum dominating set}
\class{Bounded arboricity}
\approxim{$O(\alpha)$}
\complexity{$\alpha \log n$ with high probability}
\note{The algorithm is combinatorial (does not use LPs). Improves on the previous best combinatorial algorithm, Paper~\ref{sec:LenzenW10}.}

\section{A Tight Local Algorithm for the Minimum Dominating Set Problem in Outerplanar Graphs - \cite{BonamyCGW21}}
\label{sec:BonamyCGW21}

\problem{Minimum Dominating Set}
\class{Outerplanar graphs}
\approxim{5}
\complexity{Constant}
\notes{The paper shows that the approximation ratio is tight for constant complexity on this class. The algorithm is very simple, the analysis is rather intricate.}

\section{Constant Round Distributed Domination on Graph Classes with Bounded Expansion - \cite{KublenzSV21}}
\label{sec:KublenzSV21}

\problem{Minimum dominating set}
\class{Bounded expansion}
\approxim{Constant}
\complexity{Constant}
\notes{The results improves on Paper~\ref{sec:AmiriMRS18}~\cite{AmiriMRS18} (better complexity) and on Paper~\ref{sec:CzygrinowHWW18}~\cite{CzygrinowHWW18} (larger class). The algorithm is similar to the one of Paper~\ref{sec:CzygrinowHWW18}~\cite{CzygrinowHWW18}, but the analysis is simpler and with better constants.}

\section{On Distributed Algorithms for Minimum Dominating Set problem, from theory to application - \cite{AlipourFK20}}
\label{sec:AlipourFK20}

\problem{Minimum dominating set}
\class{Planar triangle-free graphs}
\approxim{32}

\section{A LOCAL Constant Approximation Factor Algorithm for Minimum Dominating Set of Certain Planar Graphs - \cite{AlipourJ20}}
\label{sec:AlipourJ20}

\problem{Minimum dominating set}
\class{Planar graphs with no triangles and no 4-cycles}
\approxim{18}

\section{Distributed Dominating Set Approximations beyond Planar Graphs - \cite{AmiriSS19}} 
\label{sec:AmiriSS19}

\problem{Minimum dominating set}
\class{Bounded genus and a larger new class (\emph{locally embeddable graphs})}
\resultsw{
\begin{itemize}
\item Constant approximation, constant time for locally embeddable graphs
\item $(1+\epsilon)$-approximation in time $O(\log^*\!n)$ for bounded genus graphs.
\end{itemize}}
\notes{The algorithm is based on Paper~\ref{sec:LenzenPW13} \cite{LenzenPW13}, but the analysis uses new arguments.}

\section{Distributed Approximation Algorithms for the Minimum Dominating Set
in $K_h$-Minor-Free Graphs - \cite{CzygrinowHWW18}}
\label{sec:CzygrinowHWW18}
\problem{Minimum Dominating Set}
\class{$K_h$-Minor-Free Graphs}
\approxim{Constant}
\complexity{Constant}
\notes{Generalizes the results for bounded genus graphs from paper~\ref{sec:AmiriSS16}~\cite{AmiriSS16}.}

\section{Distributed Domination on Graph Classes of Bounded Expansion~-~\cite{AmiriMRS18}}\label{sec:AmiriMRS18}

\problem{$r$-(distance)-dominating set and connected $r$-dominating set}
\class{Bounded expansion}
\approxim{Constant}
\complexity{$O(\log n)$}

\section{Tight approximation bounds for dominating set on graphs of bounded arboricity - \cite{BansalU17}}
\label{sec:BansalU17}

\problem{Minimum dominating set}
\class{Bounded arboricity $\alpha$}
\approxim{$(2\alpha+1)(1+\epsilon)$}
\complexity{$\epsilon^{-4}\log^2\Delta$}
\notes{This approximation and complexity do not appear explicitly in that paper, which is centralized. But it is based on LP rounding, that can be done efficiently in CONGEST via~\cite{KuhnMW06}. See Paper~\ref{sec:Dory0I22}~\cite{Dory0I22}.}

\section{Improved distributed local approximation algorithm for minimum 2-dominating set in planar graphs - \cite{CHSWW17}}\label{sec:CHSWW17}

\problem{2-dominating set (every node covered at least twice)}
\class{Planar graphs}
\approxim{6}
\complexity{Constant}
\notes{The paper improves the approximation ratio of Paper~\ref{sec:CHSWW14} \cite{CHSWW14} from 7 from to 6, making it closer to the 4 lower bound presented in the same paper.}

\section{A Local Constant Factor {MDS} Approximation for Bounded Genus Graphs -  \cite{AmiriSS16}}\label{sec:AmiriSS16}

\problem{Minimum dominating set}
\class{Bounded-genus}
\approxim{Constant}
\complexity{$O(g)$ where $g$ is the genus}
\notes{Improves on Paper~\ref{sec:LenzenPW13} \cite{LenzenPW13}, which is for planar graph only. A different analysis of almost the same algorithm.}

\section{A local approximation algorithm for minimum dominating set problem in anonymous planar networks - \cite{Wawrzyniak15}}\label{sec:Wawrzyniak15}
\problem{Minimum dominating set}
\class{Planar}
\approxim{694}
\complexity{Constant}
\notes{Journal version of Paper~\ref{sec:Wawrzyniak13} \cite{Wawrzyniak13}. The approximation ratio is bad compared to Paper~\ref{sec:LenzenPW13} \cite{LenzenPW13} and Paper~\ref{sec:Wawrzyniak14} \cite{Wawrzyniak14}, but the model is harsher: small messages and anonymous network. (See also Paper~\ref{sec:AmiriSS16} \cite{AmiriSS16} for bounded-genus.)}

\section{A strengthened analysis of a local algorithm for the minimum dominating set problem in planar graphs - \cite{Wawrzyniak14}}\label{sec:Wawrzyniak14}
\problem{Minimum dominating set}
\class{Planar}
\approxim{52}
\complexity{Constant}
\notes{Improves the analysis of Paper~\ref{sec:LenzenPW13} \cite{LenzenPW13}, and takes the ratio from 130 down to 52.}

\section{Distributed Local Approximation of the Minimum k-Tuple Dominating Set in Planar Graphs - \cite{CHSWW14}}\label{sec:CHSWW14}

\problem{$k$-tuple minimum dominating set (every node covered at least $k$ times)}
\class{Planar}
\approxim{7 for $k=2$, $k/(k-2)$ for larger $k$}
\complexity{Constant}
\notes{For $k=2$, they also get lower bounds: $5-\epsilon$ in anonymous networks, and $4-\epsilon$ with IDs. Matching $k/(k-2)$ lower bounds for some small $k$. The 7 constant is improved to 6 in Paper~\ref{sec:CHSWW14} \cite{CHSWW17}.}

\section{Brief announcement: local approximability of minimum dominating set on planar graphs- \cite{HilkeLS13}}\label{sec:HilkeLS13}
\problem{Minimum dominating set}
\class{Planar}
\results{Lower bound on the approximation ratio for constant complexity: $(7-\epsilon)$. Improves on a ($5-\epsilon$) lower bound from Paper~\ref{sec:CHW08-disc} \cite{CHW08-disc}}

\section{Brief announcement: a local approximation algorithm for {MDS} problem in anonymous planar networks - \cite{Wawrzyniak13}}\label{sec:Wawrzyniak13}
\notes{See journal version, Paper~\ref{sec:Wawrzyniak15} \cite{Wawrzyniak15}.}

\section{Distributed minimum dominating set approximations in restricted families of graphs - \cite{LenzenPW13} }\label{sec:LenzenPW13}
\problem{Minimum dominating set}
\emph{Results:}
\begin{itemize}
\item On unit-disk graphs: ``approx $\times$ time'' $\in \Omega(\log^*n)$ (similar result but different proof in Paper~\ref{sec:CHW08-disc} \cite{CHW08-disc}). This bound is matching the upper bound.
\item On graphs of arboricity $a$: randomized $O(a^2)$-approx in time $O(\log n)$ whp. and deterministic $O(a\log(\Delta))$-approx, in time $O(\log(\Delta))$. 
\item For planar graphs, constant approx, constant time.
\end{itemize}
\notes{The paper contains a survey table for minimum dominating set. The constant approximation on planar graphs has been improved in several ways: smaller ratio (Paper~\ref{sec:Wawrzyniak14}~\cite{Wawrzyniak14}), anonymous network (Paper~\ref{sec:Wawrzyniak15} \cite{Wawrzyniak15}), and generalized to bounded genus (Paper~\ref{sec:AmiriSS16} \cite{AmiriSS16}).}

\section{Minimum Dominating Set Approximation in Graphs of Bounded Arboricity - \cite{LenzenW10}}\label{sec:LenzenW10}
\notes{See journal version, Paper~\ref{sec:LenzenPW13} \cite{LenzenPW13} (this conference version contains only the result about arboricity).}

\section{Sublogarithmic distributed MIS algorithm for sparse graphs using Nash-Williams decomposition  - \cite{BarenboimE10}}\label{sec:BarenboimE10}
\problem{Maximal independent set and coloring}
\class{Bounded arboricity}
\results{Exact algorithms. $O(\log(n)/\log\log(n))$ for MIS, $((2+\epsilon)\times a+1)$-coloring algorithm in time $O(a \log n)$}
\notes{Also, trade-off and lower bounds for coloring.}

\section{An optimal maximal independent set algorithm for bounded-independence graphs - \cite{SchneiderW10}}\label{sec:SchneiderW10}
\problem{Maximal independent set}
\class{Bounded-independance}
\complexity{$O(\log^*\!n)$}
\notes{Also, connected maximal independent set in unit-disk, maximal matching and colouring in bounded degree.}

\section{Fast Distributed Approximation Algorithm for the Maximum Matching Problem in Bounded Arboricity Graphs - \cite{CHS09}}\label{sec:CHS09}
\problem{Maximum matching}
\class{bounded arboricity}
\approxim{$1-\epsilon$}
\complexity{$O(\log^*\!n)$}
\notes{In Paper~\ref{sec:AmiriMRS18} \cite{AmiriMRS18}, the authors state that the result can be easily generalized to minor-free graphs.}

\section{What can be approximated locally?: case study: dominating sets in planar graphs - \cite{LenzenOW08}}\label{sec:LenzenOW08}
\notes{See journal version, Paper~\ref{sec:LenzenPW13} \cite{LenzenPW13}. (The proof in this conference paper is actually wrong.)}

\section{A log-star distributed maximal independent set algorithm for growth-bounded graphs -  \cite{SchneiderW08}}\label{sec:SchneiderW08}
\notes{See journal version, Paper~\ref{sec:SchneiderW10} \cite{SchneiderW10}.}

\section{Fast Distributed Approximations in Planar Graphs - \cite{CHW08-disc}}\label{sec:CHW08-disc}
\problem{Maximum independant set, maximum matching, minimum dominating set}
\approxim{$1\pm \epsilon$}
\class{Planar}
\complexity{$O(\log^*\!n)$}
\notes{Matching lower bound. Lower bound for the approximation ratio, when restricted to constant complexity (improved in Paper~\ref{sec:HilkeLS13} \cite{HilkeLS13}). Better randomized algorithm. }

\section{Distributed packing in planar graphs - \cite{CHW08-spaa}}\label{sec:CHW08-spaa}
\problem{Graph packing}
\class{Planar}
\approxim{$1-\frac{1}{\polylog n}$}
\complexity{$\polylog n$}
\notes{Generalization of the matching algorithm of \ref{sec:CHS06} \cite{CHS06}. 
Analogue of Paper~\ref{sec:CH07-disc} \cite{CH07-disc}, with planar instead of unit-disk.}

\section{A randomized distributed algorithm for the maximal independent set problem in growth-bounded graphs - \cite{GfellerV07}}\label{sec:GfellerV07}
\problem{Maximal independent set}
\class{Bounded growth}
\complexity{$O(\log\log n \times \log^*\!n)$ whp}

\section{Distributed Approximation Algorithms for Weighted Problems in Minor-Closed Families - \cite{CH07-cocoon}}\label{sec:CH07-cocoon}
\problem{Weighted maximum matching, weighted minimum dominating set}
\class{Minor-free}
\approxim{$1-\frac{1}{\polylog n}$}
\complexity{$\polylog n$}
\notes{Introduces a new clustering, to generalize the results of Paper~\ref{sec:CH06-esa} \cite{CH06-esa}. A lower bound for weighted connected minimum dominating set.}

\section{Distributed Approximations for Packing in Unit-Disk Graphs - \cite{CH07-disc}}\label{sec:CH07-disc}
\problem{Graph packing}
\class{Unit-disk}
\approxim{$1-\frac{1}{\polylog n}$}
\complexity{$\polylog n$}
\notes{Extension of Paper~\ref{sec:CH06-disc} \cite{CH06-disc} to a more general class of graphs. Analogue of Paper~\ref{sec:CHW08-spaa} \cite{CHW08-spaa}.}

\section{Distributed Almost Exact Approximations for Minor-Closed Families -  \cite{CH06-esa}}\label{sec:CH06-esa}
\problem{Minimum dominating set and connected minimum dominating set}
\class{Minor-free}
\approxim{$1-\frac{1}{\polylog n}$}
\complexity{$\polylog n$}
\notes{Extends Paper~\ref{sec:CHS06} \cite{CHS06}: no additional assumption, and larger class. Generalizes also the clustering of Paper~\ref{sec:CH06-jda} \cite{CH06-jda}. The minimum dominating set part is generalized to a weighted version in Paper~\ref{sec:CH07-cocoon} \cite{CH07-cocoon}.}

\section{Distributed Approximation Algorithms for Planar Graphs  - \cite{CHS06}}\label{sec:CHS06}
\problem{Maximum matching and minimum dominating set}
\class{planar and planar with some forbidden minor}
\approxim{$1-\frac{1}{\polylog n}$}
\complexity{$\polylog n$}
\notes{Completely generalized in later works: minimum dominating set generalized in Paper~\ref{sec:CH06-esa} \cite{CH06-esa}, and matching generalized to packing in Paper~\ref{sec:CHW08-spaa} \cite{CHW08-spaa}.}

\section{Distributed algorithms for weighted problems in sparse graphs~-~\cite{CH06-jda}}\label{sec:CH06-jda}
\problem{Weighted minimum dominating set, weighted matching, weighted maximum independent set}
\class{Trees, except for independent set for which it is planar graphs}
\approxim{$1-\frac{1}{\polylog n}$}
\complexity{$\polylog n$}
\notes{Clustering generalized in Paper~\ref{sec:CH06-esa} \cite{CH06-esa}. Results for weighted minimum dominating set and weighted matching are generalized to minor-free graphs in Paper~\ref{sec:CH07-cocoon} \cite{CH07-cocoon}.}

\section{Distributed Approximation Algorithms in Unit-Disk Graphs - \cite{CH06-disc} }\label{sec:CH06-disc}
\notes{Completely generalized in Paper~\ref{sec:CH07-disc} \cite{CH07-disc}.}

\section{Fast Deterministic Distributed Maximal Independent Set Computation on Growth-Bounded Graphs - \cite{KuhnMNW05-disc}}\label{sec:KuhnMNW05-disc}
\problem{Maximal independent set}
\class{Bounded-growth}
\complexity{$O(\log^*\!n\times\log\Delta)$}

\section{Local approximation schemes for ad hoc and sensor networks~-~ \cite{KuhnNMW05-dialm}}\label{sec:KuhnNMW05-dialm}
\problem{Minimum dominating set and maximum independent set}
\class{Unit-disk and polynomially bounded growth}
\approxim{$1+\epsilon$}
\complexity{$T + \frac{\log^*n}{\epsilon^{O(1)}}$, where $T$ is the time to compute an MIS in the class. That is a PTAS}

\section{On the locality of bounded growth - \cite{KuhnMW05}}\label{sec:KuhnMW05}
\problem{Arbitrary covering and packing LPs (the fractional versions), and decompositions}
\class{Unit-disk and bounded growth}
\notes{The decomposition holds in unit-disk graphs with doubling metric, and gives for example constant approximation for minimum dominating set in such graphs in time $O(\log^*\!n)$.}

\section{Distributed Algorithm for Better Approximation of the Maximum Matching - \cite{CH03}}\label{sec:CH03}
\problem{Maximum matching}
\class{Graphs without odd cycles of length 3, 5, ..., $2k-1$}
\approxim{$1-\frac{1}{k+1}$}
\complexity{$\polylog n$}

\newpage{}

\DeclareUrlCommand{\Doi}{\urlstyle{same}}
\renewcommand{\doi}[1]{\href{http://dx.doi.org/#1}{\footnotesize\sf doi:\Doi{#1}}}
\bibliography{distributed_approx.bib}{}
\bibliographystyle{plainnat}

\end{document}